\documentclass[11pt,twoside]{article}
\usepackage{asp2010}

\resetcounters

\begin{document}

\title{Chemical evolution of the Galactic disk(s)}
\author{Thomas Bensby and Sofia Feltzing,
\affil{Lund Observatory, Box 43, SE-221\,00 Lund , Sweden}}

\begin{abstract}
We highlight some results from our high-resolution spectroscopic
elemental abundance survey  of 
F and G dwarf stars in the solar neighbourhood.  
\end{abstract}

\section{Introduction}

To investigate the chemical and 
kinematical properties of the Galactic disk, and the thick disk
in particular, we have undertaken a spectroscopic survey of 703 
kinematically selected F and G 
dwarf stars in the Solar neighbourhood. Based on high resolution 
($R=45\,000$ to 110\,000) and high signal-to-noise 
($S/N\approx 150$ to 300) spectra for all stars we determined
detailed elemental abundances for O, Na, Mg, Al, Si, Ca, Ti, Cr, Fe, Ni, Zn, Y,
Ba, and stellar ages from isochrones. Including the results for the
first 102 stars of the sample presented in \cite{bensby2003,bensby2005},
our main findings include: 
(i) at a given metallicity, the thick disk abundance trends are 
more $\alpha$-enhanced than those of the thin disk; 
(ii) the metal-rich limit of the thick disk reaches at least 
solar metallicities \citep{bensby2007letter2}; 
(iii) the metal-poor limit of the thin disk is around 
$\rm [Fe/H]\approx-0.8$; (iv) the thick disk shows an age-metallicity gradient;
(v) the thin disk does {\it not} show an age-metallicity gradient; (vi) the most
metal-rich thick disk stars at $\rm [Fe/H]\approx 0$ are significantly older than the
most metal-poor thin disk stars at $\rm [Fe/H]\approx -0.7$; (vii) based on our elemental 
abundances we find that kinematical
criteria produce thin and thick disk stellar samples that are biased in the sense
that stars from the low-velocity tail of the thick disk are classified as thin disk
stars, and stars from the high-velocity tail of the thin disk are classified 
as thick disk stars; (viii) age criteria appears to produce thin and thick
disk stellar samples with less contamination. These points were recently discussed
in \cite{bensby2011_lgb}. Based on the current sample we have also found that 
the Hercules stream is likely to be of dynamical origin and that its stars just is
a mix of thin and thick disk stars \citep{bensby2007letter}.
In this proceeding we will discuss new findings about the variation of 
the elemental abundance ratios in the Galactic disk with Galactocentric 
distance.

\section{Variation with Galactocentric radius}

\begin{figure}
\resizebox{\hsize}{!}{
\includegraphics[bb=-20 162 630 490,clip]{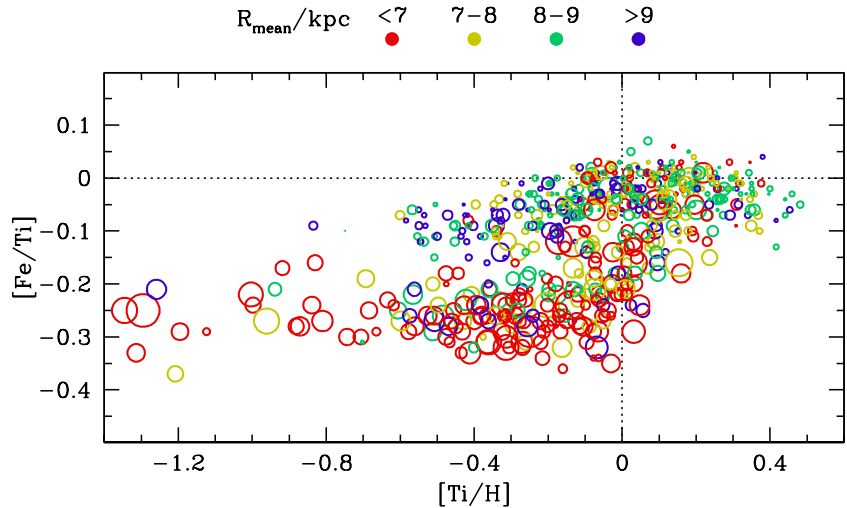}}
\caption{[Fe/Ti] - [Ti/H] abundance plot. 
The color-coding refers to the mean Galactocentric radius of the
orbits of the stars and is defined at the top of the figure. The sizes
of the circles have been scaled by the ages of the stars (larger = older).
\label{fig:feti}
                    }
   \end{figure}

Fig.~\ref{fig:feti} shows the [Fe/Ti]-[Ti/H] abundance plot for the
full sample. The stars have been colour-coded based on their orbits
mean distance from the Galactic centre ($R_{\rm mean}$),
as well as size-coded based on their estimated ages. 
A  majority of the stars with 
$R_{\rm mean}<7$\,kpc are $\alpha$-enhanced and have high ages (i.e. red 
and big circles). The opposite is observed for the stars with 
$R_{\rm mean}>9$\,kpc which tend to be less $\alpha$-enhanced and 
have lower ages (i.e. blue and small circles). 

If $R_{\rm mean}$ can be associated with the approximate birthplace of a
star, and if a main characteristic of the Galactic thick
disk is that its stars mainly are old and $\alpha$-enhanced, the above finding 
indicates that the thick disk stars in the solar neighbourhood, 
should mainly originate from the inner disk region. Younger and less $\alpha$-enhanced
(thin disk) stars on the other hand are more prone to originate from the 
outer disk region. This result agrees well
with the study of red giants located in the inner and outer disk by 
\cite{bensby2010letter,bensby2011letter} where it was found
that the stars with thick disk abundance patterns 
appears extremely sparse in the outer disk, even at large distances from the 
Galactic plane. In the inner disk region on the other hand,
both thin and thick disk abundance patterns were found.
A possible explanation is that the scale-length for the thick disk
is significantly shorter than for the thin disk \citep{bensby2011letter}.
What this means for the existence of an outer thick disk, and the dichotomy
of the Galactic disk as seen in the solar neighbourhood, will be
further investigated in an upcoming paper.

\acknowledgements T.B. was funded by grant No. 621-2009-3911 from The 
Swedish Research Council. S.F. was partly funded by the Swedish Royal 
Academy of Sciences and partly by grant No. 2008-4095 from The Swedish 
Research Council.
\bibliographystyle{asp2010}
\bibliography{referenser}

\end{document}